\documentclass[aip,reprint]{revtex4-1}
\usepackage{graphicx}
\usepackage{appendix}
\usepackage{braket}
\usepackage{rotating}
\usepackage{amssymb}
\usepackage{amsmath}
\usepackage{txfonts}
\usepackage{bm}
\usepackage{color}
\usepackage{multirow}
\usepackage{booktabs}
\usepackage{dcolumn}
\usepackage{verbatim}

\begin{document}

\title{Thermal quasi-particle theory}

\author{So Hirata}
\email{sohirata@illinois.edu}
\affiliation{Department of Chemistry, University of Illinois at Urbana-Champaign, Urbana, Illinois 61801, USA}

\date{\today}

\begin{abstract}
The widely used thermal Hartree--Fock (HF) theory is generalized to include the effect of electron correlation while maintaining its quasi-independent-particle framework.
An electron-correlated internal energy (or grand potential) is postulated in consultation with the second-order finite-temperature many-body perturbation theory (MBPT), 
which then dictates the corresponding thermal orbital (quasi-particle) 
energies in such a way that all fundamental thermodynamic relations are obeyed. 
The associated density matrix is of a one-electron type, whose diagonal elements take the form of the Fermi--Dirac distribution functions, 
when the grand potential is minimized. The formulas for the entropy and chemical potential are unchanged from 
those of Fermi--Dirac or thermal HF theory. 
The theory thus constitutes a finite-temperature extension of the second-order Dyson self-energy of one-particle many-body Green's function theory and can be viewed
as a second-order, diagonal, frequency-independent, thermal inverse Dyson equation. At low temperature, the theory approaches finite-temperature MBPT of the same order,
but it may outperform the latter at intermediate temperature by including additional electron-correlation effects through orbital energies. 
A physical meaning of these thermal orbital energies is proposed (encompassing that of thermal HF orbital energies, which has been elusive) as a finite-temperature version of Janak's theorem.
\end{abstract}

\maketitle 

\section{Introduction} 

Thermal Hartree--Fock (HF) theory\cite{Husimi1940,Mermin63,Gu24} is a curious ansatz. It uses an auxiliary one-electron Hamiltonian to define its one-electron density matrix, whereas a true density matrix for thermodynamics 
governed by  an interacting Hamiltonian should be much higher ranked.\cite{Farid2000} 
Its diagonal elements, as variational parameters, ultimately
become the Fermi--Dirac distribution functions of the non-interacting Fermi--Dirac theory, when the grand potential is minimized. 
The state energies defining this grand potential are, however, evaluated with the exact Hamiltonian containing two-electron interactions. Stipulated in this way, 
the theory does not seem to correspond to a single, consistent grand partition function. 

Nevertheless, the thermodynamic functions of this theory are shown\cite{Argyres1974,Gu24} to obey 
all fundamental thermodynamic relations. They also correctly reduce\cite{Gu24}
to the zero-temperature HF theory.\cite{szabo,shavitt} 
They are widely used in applications to, e.g., energy bands in a solid 
where their temperature-dependent orbital energies are often invoked, somewhat unquestioningly, in explaining
metal-insulator transitions,\cite{Hermes2015} etc. This is in spite of the facts that 
their physical meaning is still obscure\cite{Pain} and that state energies in quantum mechanics are supposed to be constant of temperature.

In this article, inspired by the immense success of thermal HF theory, we generalize this ansatz to include the effects of electron correlation, while keeping to its quasi-independent-particle framework
and grand canonical ensemble. Such a theory, which can provide electron-correlated (i.e., quasi-particle) energy bands of semiconductors at finite temperatures in addition to their 
thermodynamic functions, may prove even more useful
than thermal HF theory or thermal density-functional theory (DFT). 
Our strategy of postulating what we call {\it thermal quasi-particle theory} is as follows: 

We first assume that every state consists of
non-interacting quasi-particles, each occupying a well-defined orbital with an electron-correlated orbital energy. 
An internal energy (or grand potential) is postulated by consulting with finite-temperature many-body perturbation theory (MBPT).\cite{HirataJha,HirataJhaJCP2020,Hirata2021}
Its definition dictates the forms of the thermal quasi-particle energies (i.e., the electron-correlated thermal orbital energies) such that all fundamental thermodynamic relations are obeyed. 
The corresponding density matrix is then of a one-electron type, whose diagonal elements take the form of the Fermi--Dirac distribution functions, 
when the grand potential is minimized. The formulas for the 
entropy and average number of electrons (determining the chemical potential) are unchanged from the corresponding formulas of Fermi--Dirac or thermal HF theory. 
The thermal quasi-particle energies thus defined are  a finite-temperature generalization\cite{matsubara,luttingerward,march,Kadanoff_book} of 
the Dyson self-energies of one-particle many-body Green's function (MBGF) theory.\cite{linderbergohrn,paldus,cederbaumacp,simonsrev,ohrnborn,jorgensensimons,oddershede,GW1,ortiz_aqc,GW2,Wire,Hirata2017,Hirata_PRA2024}

Being based on perturbation theory, thermal quasi-particle theory forms a hierarchy of size-consistent\cite{HirataTCA} approximations with increasing accuracy and complexity.
Fermi--Dirac and thermal HF theories can be viewed, respectively, as the zeroth- and first-order instances of this hierarchy. Here, we 
introduce a second-order thermal quasi-particle theory, which is the leading order in describing electron correlation. It is based on the second-order grand potential or internal energy
of finite-temperature MBPT.\cite{HirataJhaJCP2020,Hirata2021} 
We propose the expression for the corresponding second-order thermal self-energies, which obey all thermodynamic relations and ensure the variationality of the grand potential.
The theory thus constitutes a second-order, diagonal, frequency-independent, thermal inverse Dyson equation.\cite{Hirata2017} 
It reduces to the second-order MBPT for internal energy and to the second-order MBGF (in the diagonal and frequency-independent approximation) for orbital energies
at zero temperature.   

A comparison of thermal quasi-particle theories with finite-temperature MBPT suggests that the former may include higher-order perturbation 
corrections through correlation-corrected thermal orbital energies and outperform the latter at intermediate temperature. 
We also propose a physical meaning of these thermal orbital energies in the form of a finite-temperature version of Janak's theorem.\cite{Janak1978}
It encompasses the physical meaning of thermal HF orbital energies, which has been elusive.\cite{Pain,Hirata2021} 

\section{Thermal quasi-particle theory} 
\subsection{Zeroth order\label{sec:FD}}

Fermi--Dirac theory is the zeroth-order thermal quasi-particle theory [thermal QP(0) theory]. It is reviewed briefly to outline the general framework
of the hierarchical approximations; it is discussed more fully in Ref.\ \onlinecite{Gu24}. 

Its internal energy $U^{(0)}$ and entropy $S^{(0)}$ at the inverse temperature $\beta = (k_\text{B}T)^{-1}$ are given by
\begin{eqnarray}
U^{(0)} &\equiv& \langle E^{(0)} \rangle =  \sum_p \epsilon_p^{(0)} f_p^-, \label{FD:U} \\
S^{(0)} &\equiv& -k_{\text{B}}  \left( f_p^- \ln f_p^- + f_p^+ \ln f_p^+  \right) \label{FD:S}
\end{eqnarray}
with the Fermi--Dirac distribution functions,
\begin{eqnarray}
f_p^- &=& \frac{1}{1+e^{\beta (\epsilon_{p}^{(0)} - \mu^{(0)})}}, \label{FD:minus} \\
f_p^+ &=&1 - f_p^- = \frac{e^{\beta (\epsilon_{p}^{(0)}-\mu^{(0)})}}{1+e^{\beta (\epsilon_{p}^{(0)}-\mu^{(0)})}}. \label{FD:plus}
\end{eqnarray}
Here, $\epsilon_p^{(0)}$ is the energy of the $p$th spinorbital of a reference wave function, and $\mu^{(0)}$ is the chemical potential
determined by the electroneutrality condition,
\begin{eqnarray}
\bar{N} &=& \sum_p f_p^-, \label{FD:mu} 
\end{eqnarray}
where $\bar{N}$ is the average number of electrons that cancel nuclear charges exactly, and the summations are always taken over all spinorbitals. 

The $\langle E^{(n)} \rangle$ in Eq.\ (\ref{FD:U}) is a zeroth-order thermal average of state energies, i.e.,
\begin{eqnarray}
\langle E^{(n)} \rangle &\equiv& \frac{\sum_I E_I^{(n)} e^{-\beta(E_I^{(0)} - \mu^{(0)} N_I)}}{\sum_I e^{-\beta(E_I^{(0)} - \mu^{(0)} N_I)}},
\end{eqnarray}
where $E_I^{(n)}$ is the $n$th-order perturbation correction to the energy of the $I$th Slater-determinant 
state according to Hirschfelder--Certain degenerate perturbation theory,\cite{Hirschfelder} and 
$E_I^{(0)} = \sum \epsilon_i^{(0)}$  ($i$ labels spinorbitals occupied in the $I$th state). Sum-over-states expressions of the thermal averages 
can be reduced to sum-over-orbitals ones either by combinatorial logic\cite{HirataJha,HirataJhaJCP2020} or
 by normal-ordered second quantization at finite temperature.\cite{Hirata2021}

The grand potential $\Omega^{(0)}$ is then given by
\begin{eqnarray}
\Omega^{(0)} &\equiv& U^{(0)} - \mu^{(0)} \bar{N} - TS^{(0)}  \label{FD:relation0}  \\
&=& \sum_p  \left( \epsilon_p^{(0)}  -\mu^{(0)} \right) f_p^- 
+ \frac{1}{\beta} \sum_p \left( f_p^- \ln f_p^- + f_p^+ \ln f_p^+  \right). \nonumber\\ \label{FD:Omega}
\end{eqnarray}

As explicitly shown in Ref.\ \onlinecite{Gu24}, they satisfy the thermodynamic relations such as Eq.\ (\ref{FD:relation0}) and
\begin{eqnarray}
-\frac{\partial\Omega^{(0)}}{\partial \mu^{(0)}}&=&\bar{N}  , \label{FD:relation1}\\
 -\frac{\partial \Omega^{(0)}}{\partial T}&=& S^{(0)}. \label{FD:relation2} 
\end{eqnarray}

The Fermi--Dirac distribution function [Eq.\ (\ref{FD:minus})] is a diagonal element of the one-electron density matrix that minimizes 
the grand potential. Therefore,
\begin{eqnarray}
\frac{\partial \Omega^{(0)}}{\partial f_p^-} &=& 0, \label{FD:relation3}
\end{eqnarray}
which is also explicitly verifiable.\cite{Gu24}

Fermi--Dirac theory is an exact theory for thermodynamics of a system governed by an independent-particle Hamiltonian. 

\subsection{First order}

Thermal HF theory\cite{Husimi1940,Mermin63} constitutes the first-order thermal quasi-particle theory [thermal QP(1) theory]. It is also
fully discussed in Ref.\ \onlinecite{Gu24}. It accommodates the exact Hamiltonian with two-electron interactions within
a quasi-independent-particle framework. 

Its internal energy $U^\text{HF}$ is postulated by an intuitively natural finite-temperature generalization of the zero-temperature HF energy,\cite{Hirata2021} i.e.,
\begin{eqnarray}
U^\text{HF} &\equiv& \langle E^{(0)} \rangle + \langle E^{(1)} \rangle \\
&=& \sum_p \epsilon_p^\text{HF} f_p^- - \frac{1}{2} \sum_{p,q} \langle pq || pq \rangle f_p^- f_q^-   \label{HF:U} 
\end{eqnarray}
with
\begin{eqnarray}
\epsilon_{p}^\text{HF} &\equiv& h_{pp}+ \sum_r \langle pr || pr \rangle f_r^-, \label{HF:epsilondiag}
\end{eqnarray}
where $h_{pq}$ is the one-electron (``core'') part of the Hamiltonian matrix element,\cite{szabo} $\langle pq || rs \rangle$ is an anti-symmetrized two-electron integral,\cite{szabo,shavitt} and 
$\langle E^{(1)} \rangle$ has been reduced in Eqs.\ (45) and (46) of Ref.\ \onlinecite{HirataJhaJCP2020}. 
Spinorbitals labeled by $p$ and $q$ 
are the ones that bring the matrix of thermal HF orbital energies $\bm{\epsilon}^\text{HF}$ into a diagonal form, i.e., 
\begin{eqnarray}
\epsilon_{pq}^\text{HF} &=& h_{pq}+ \sum_r \langle pr || qr \rangle f_r^-  = \delta_{pq}\epsilon_p^\text{HF} 
, \label{HF:epsilon}
\end{eqnarray}
where $\delta_{pq}$ is Kronecker's delta. 

The Fermi--Dirac distribution functions, 
\begin{eqnarray}
f_p^- &=& \frac{1}{1+e^{\beta (\epsilon_{p}^\text{HF} - \mu^\text{HF})}}, \label{HF:minus} \\
f_p^+ &=&1 - f_p^- = \frac{e^{\beta (\epsilon_{p}^\text{HF}-\mu^\text{HF})}}{1+e^{\beta (\epsilon_{p}^\text{HF}-\mu^\text{HF})}}, \label{HF:plus}
\end{eqnarray}
are now defined with $\epsilon_p^\text{HF}$ and $\mu^\text{HF}$. The latter is determined by the same electroneutrality condition,
\begin{eqnarray}
\bar{N} &=& \sum_p f_p^-. \label{HF:mu}
\end{eqnarray}

The entropy formula is unchanged from that of Fermi--Dirac theory [Eq.\ (\ref{FD:S})],
\begin{eqnarray}
S^\text{HF} &\equiv& -k_{\text{B}} \sum_p \left(  f_p^- \ln f_p^-+ f_p^+ \ln f_p^+ \right) . \label{HF:S}
\end{eqnarray}
The grand potential is therefore given by
\begin{eqnarray}
\Omega^\text{HF} &\equiv& U^\text{HF} - \mu^\text{HF} \bar{N} - TS^\text{HF}  \label{HF:relation0} \\
&=& \sum_p  \left( \epsilon_p^\text{HF} -  \mu^\text{HF} \right) f_p^- - \frac{1}{2} \sum_{p,q} \langle pq || pq \rangle f_p^- f_q^-   \nonumber\\
&& +\frac{1}{\beta} \sum_p \left(  f_p^- \ln f_p^-+ f_p^+ \ln f_p^+ \right).  \label{HF:Omega}
\end{eqnarray}

These thermodynamic functions together satisfy\cite{Gu24}
the thermodynamic relations such as Eq.\ (\ref{HF:relation0}) by construction as well as 
\begin{eqnarray}
-\frac{\partial\Omega^\text{HF}}{\partial \mu^\text{HF}} &=&\bar{N} , \label{HF:relation1}\\
-\frac{\partial \Omega^\text{HF}}{\partial T} &=& S^\text{HF} , \label{HF:relation2} \\
\frac{\partial \Omega^\text{HF}}{\partial f_p^-} &=& 0. \label{HF:relation3} 
\end{eqnarray}
The last identity underscores that $\Omega^\text{HF}$ is minimized\cite{Mermin63,Gu24}  by the orbitals that diagonalize $\bm{\epsilon}^\text{HF}$ [Eq.\ (\ref{HF:epsilon})]  and by the one-particle density matrix whose eigenvalues are
the Fermi--Dirac distribution functions of Eq.\ (\ref{HF:minus}).

\subsection{Second order}

We postulate a second-order thermal quasi-particle theory [thermal QP(2) theory] by its internal energy,
\begin{eqnarray}
U^\text{QP(2)} &\equiv& \langle E^{(0)} \rangle +  \langle E^{(1)} \rangle +  \langle E^{(2)} \rangle .
\end{eqnarray}
The last term $\langle E^{(2)} \rangle$ is the thermal average of second-order Hirschfelder--Certain degenerate perturbation energies\cite{Hirschfelder}  and has been reduced in Eqs.\ (C7) and (C8) of Ref.\ \onlinecite{HirataJhaJCP2020} to 
the following sum-over-orbitals formula:
\begin{eqnarray}
\langle E^{(2)} \rangle &=&
\sum_{p,q}^{\text{denom.}\neq0} \frac{F_{qp} F_{pq} }{\epsilon_p^{(0)} - \epsilon_q^{(0)}} f_p^- f_q^+ \nonumber\\ 
&& + 
\frac{1}{4} \sum_{p,q,r,s}^{\text{denom.}\neq0} \frac{\langle pq||rs \rangle \langle rs || pq \rangle}{\epsilon_p^{(0)} + \epsilon_q^{(0)} - \epsilon_r^{(0)} - \epsilon_s^{(0)}} f_p^-f_q^-f_r^+f_s^+  \nonumber\\ \label{E2}
\end{eqnarray}
with
\begin{eqnarray}
F_{pq} \equiv \epsilon_{pq}^\text{HF} - \delta_{pq} \epsilon_p^{(0)},
\end{eqnarray}
where ``denom.$\neq0$'' restricts the summation to over those indexes whose 
denominator is nonzero.\cite{HirataJhaJCP2020,Hirata2021} 

It is a natural finite-temperature generalization of second-order many-body perturbation energy,\cite{bartlett_arpc,shavitt} but is considerably simpler than
the full second-order correction  to the internal energy $U^{(2)}$ of finite-temperature MBPT [Eq.\ (74) of Ref.\ \onlinecite{HirataJhaJCP2020}].
The simplification is at least  partly justified by the equally simpler treatments of the chemical-potential and entropy contributions to adhere to the quasi-independent-particle framework 
that the theory adopts.

\begin{figure}
  \includegraphics[scale=0.4]{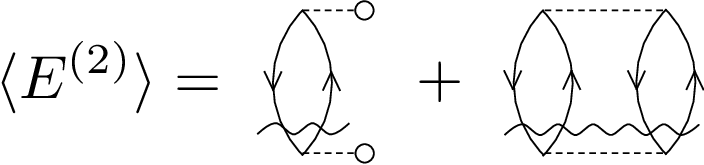}
\caption{Shavitt--Bartlett-style diagrams\cite{shavitt,Hirata2021} of  $\langle E^{(2)} \rangle$ of Eq.\ (\ref{E2}). 
Each downgoing (upgoing) edge represents $f_p^-$ ($f_p^+$), a dashed line is an anti-symmetrized two-electron integral,
a dashed line with a circle denotes $F_{pq}$, and a wiggly (resolvent) line is a denominator factor. }
\label{fig:E2}
\end{figure}

The first term of $\langle E^{(2)} \rangle$ [Eq.\ (\ref{E2})] is the so-called non-HF term (or more precisely, non-thermal-HF term, in this case),\cite{shavitt} which is zero when thermal HF theory is used as the reference (where $F_{pq} = 0$). 
The diagrammatic representation\cite{Hirata2021} of $\langle E^{(2)} \rangle$ is given in Fig.\ \ref{fig:E2}.

We thus have
\begin{eqnarray}
U^\text{QP(2)} &\equiv&  \sum_p \epsilon_p^\text{HF} f_p^- - \frac{1}{2} \sum_{p,q} \langle pq || pq \rangle f_p^- f_q^-  
\nonumber\\&& 
+ \sum_{p,q}^{\text{denom.}\neq0} \frac{F_{qp} F_{pq} }{\epsilon_p^{(0)} - \epsilon_q^{(0)}} f_p^- f_q^+ \nonumber\\ 
&& + \frac{1}{4} \sum_{p,q,r,s}^{\text{denom.}\neq0} \frac{\langle pq||rs \rangle \langle rs || pq \rangle}{\epsilon_p^{(0)} + \epsilon_q^{(0)} - \epsilon_r^{(0)} - \epsilon_s^{(0)}} f_p^-f_q^-f_r^+f_s^+. \nonumber\\ \label{QP:U} 
\end{eqnarray}
The entropy formula retains the one-particle picture and takes the form, 
\begin{eqnarray}
S^\text{QP(2)} &\equiv& -k_{\text{B}} \sum_p \left(  f_p^- \ln f_p^-+ f_p^+ \ln f_p^+ \right), \label{QP:S}
\end{eqnarray}
which differs materially from the $S^{(0)}+S^{(1)}+S^{(2)}$ of Refs.\ \onlinecite{HirataJhaJCP2020,Hirata2021}, but is 
more in line with thermal HF or Fermi--Dirac theory. 
The corresponding grand potential is then defined by
\begin{eqnarray}
\Omega^\text{QP(2)} &\equiv& U^\text{QP(2)} - \mu^\text{QP(2)} \bar{N} - TS^\text{QP(2)}  \label{QP:relation0} \\
&=& \sum_p \left(  \epsilon_p^\text{HF} -\mu^\text{QP(2)} \right) f_p^- - \frac{1}{2} \sum_{p,q} \langle pq || pq \rangle f_p^- f_q^-  \nonumber\\
&& +\frac{1}{\beta} \sum_p \left(  f_p^- \ln f_p^-+ f_p^+ \ln f_p^+ \right) + \langle E^{(2)} \rangle , \label{QP:Omeganew}
\end{eqnarray}
where $\mu^\text{QP(2)}$ is the chemical potential, which is determined by the same electroneutrality condition of Fermi--Dirac theory, which reads
\begin{eqnarray}
\bar{N} &=& \sum_p f_p^-. \label{QP:mu}
\end{eqnarray}

Spinorbitals labeled by $p$ and $q$ are those of a reference theory, which is typically but not limited to 
zero-temperature HF theory or thermal HF theory at the same $T$. 
No orbital rotation by matrix diagonalization is performed in this ansatz, but $\Omega^\text{QP(2)}$ is still minimized with respect to 
the diagonal elements of the one-electron density matrix, which are none other than the $f_p^-$. 

These $f_p^\mp$ take the form of the Fermi--Dirac distribution functions, when $\Omega^\text{QP(2)}$ is minimized by varying $f_p^-$ (see below for a proof).\cite{Gu24} 
They are defined with the second-order quasi-particle energy $\epsilon_p^\text{QP(2)}$ and chemical potential $\mu^\text{QP(2)}$, i.e.,
\begin{eqnarray}
f_p^- &=& \frac{1}{1+e^{\beta (\epsilon_{p}^\text{QP(2)} - \mu^\text{QP(2)})}}, \label{QP:minus} \\
f_p^+ &=&1 - f_p^- = \frac{e^{\beta (\epsilon_{p}^\text{QP(2)}-\mu^\text{QP(2)})}}{1+e^{\beta (\epsilon_{p}^\text{QP(2)}-\mu^\text{QP(2)})}} \label{QP:plus}
\end{eqnarray}
with $\epsilon_p^\text{QP(2)}$ given by
\begin{eqnarray}
\epsilon_p^\text{QP(2)} &=&\epsilon_p^\text{HF} + \Sigma_{pp}^{(2)}. \label{QP:epsilon}
\end{eqnarray}

\begin{figure}
  \includegraphics[scale=0.4]{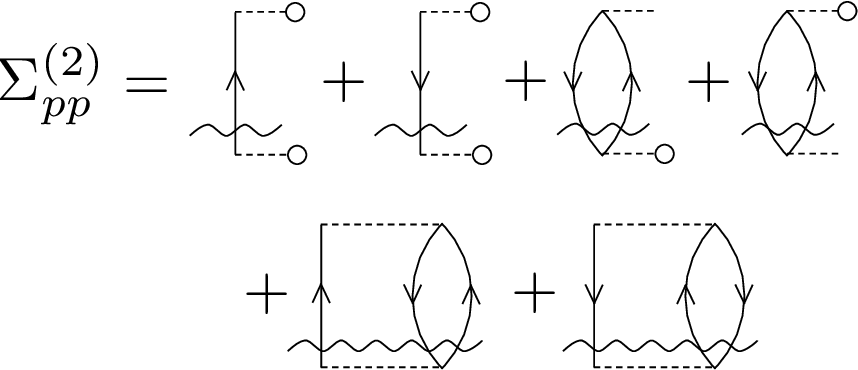}
\caption{Second-order thermal Dyson self-energy diagrams in the zero-temperature reference [Eq.\ (\ref{QP:sigma2_e0})], which are 
obtained by deleting an edge in each of the internal energy diagrams in Fig.\ \ref{fig:E2}.
See also the caption of Fig.\ \ref{fig:E2} as well as Refs.\ \onlinecite{Hirata2021,Hirata2017}.}
\label{fig:Sigma2_e0}
\end{figure}

Here, $\Sigma_{pp}^{(2)}$ is a finite-temperature analogue of the second-order Dyson self-energy in the Feynman--Dyson perturbation expansion of MBGF.\cite{Hirata2017} 
In a zero-temperature reference, where $\epsilon_p^{(0)}$ is constant of $T$, it takes the form of
\begin{eqnarray}
\Sigma_{pp}^{(2)} &=&
\sum_{q}^{\text{denom.}\neq0} \frac{ F_{qp}F_{pq} }{\epsilon_p^{(0)} - \epsilon_q^{(0)}} f_q^+ 
 - \sum_{q}^{\text{denom.}\neq0} \frac{F_{pq}F_{qp} }{\epsilon_q^{(0)} - \epsilon_p^{(0)}} f_q^- \nonumber\\ 
&& + \sum_{q,r}^{\text{denom.}\neq0} \frac{\langle qp || rp \rangle F_{rq} }{\epsilon_r^{(0)} - \epsilon_q^{(0)}} f_r^- f_q^+ 
\nonumber\\&& + \sum_{q,r}^{\text{denom.}\neq0} \frac{ F_{qr} \langle rp ||qp \rangle }{\epsilon_r^{(0)} - \epsilon_q^{(0)}} f_r^- f_q^+ \nonumber\\
&& +  \frac{1}{2} \sum_{q,r,s}^{\text{denom.}\neq0} \frac{\langle pq||rs \rangle \langle rs || pq \rangle}{\epsilon_p^{(0)} + \epsilon_q^{(0)} - \epsilon_r^{(0)} - \epsilon_s^{(0)}}  f_q^-f_r^+f_s^+
\nonumber\\
&& - \frac{1}{2} \sum_{q,r,s}^{\text{denom.}\neq0} \frac{\langle rs || pq \rangle \langle pq||rs \rangle}{\epsilon_r^{(0)} + \epsilon_s^{(0)} - \epsilon_p^{(0)} - \epsilon_q^{(0)}}  f_q^+f_r^-f_s^-, 
 \label{QP:sigma2_e0}
\end{eqnarray}
which is obtained by demanding all thermodynamic relations be obeyed (see below for a derivation). A diagrammatic representation of $\Sigma_{pp}^{(2)}$ is given in Fig.\ \ref{fig:Sigma2_e0}. 
 
The first two terms of the above equation (or the first two diagrams in Fig.\ \ref{fig:Sigma2_e0}) are the so-called semi-reducible diagrams of $\Delta$MP2.\cite{deltamp} 
Not only do they account for the non-HF-reference contributions,\cite{shavitt} but they 
correct the errors arising from the diagonal approximation to the self-energy implicit in the $\Delta$MP$n$ ansatz.\cite{deltamp} Thanks to these diagrams, which are illegal in the Feynman--Dyson 
perturbation expansion of MBGF,\cite{Hirata2017} 
$\Delta$MP$n$ is convergent at the exactness (and, in fact, more reliably so\cite{Hirata_PRA2024}  than Feynman--Dyson MBGF) while using the diagonal and frequency-independent approximations throughout the 
perturbation orders. 
In this sense, the thermal self-energy thus defined may be more aptly viewed as a finite-temperature analogue of 
$\Delta$MP$n$ (Ref.\ \onlinecite{deltamp}) than of Feynman--Dyson MBGF (Ref.\ \onlinecite{Hirata2017}). 

The third and fourth terms (diagrams) are non-HF terms\cite{shavitt} and are also related to the energy-independent diagrams.\cite{purvis_cpl,SchirmerAngonoa} 

When a zero-temperature HF reference is used, in the $T = 0$ limit, the first four terms
vanish because $F_{pq}=0$, leaving only the last two terms as the zero-temperature second-order self-energy in the diagonal, frequency-independent approximation.
Their diagrams (the last two in Fig.\ \ref{fig:Sigma2_e0}) are, therefore, isomorphic to the two second-order self-energy diagrams (e.g., Fig.\ 4 of Ref.\ \onlinecite{Hirata2017}) of zero-temperature MBGF, although 
their scope is different (the former are for $T \geq 0$, while the latter are only for $T=0$). 

With these definitions, the thermodynamic relation of Eq.\ (\ref{QP:relation0}) is satisfied by construction. 
The next one,
\begin{eqnarray}
-\frac{\partial\Omega^\text{QP(2)}}{\partial \mu^\text{QP(2)}} &=& \bar{N}  , \label{QP:relation1}
\end{eqnarray}
is also obeyed, which can be confirmed by explicit evaluation of the derivative.
\begin{eqnarray}
- \frac{\partial \Omega^\text{QP(2)}}{\partial \mu^\text{QP(2)}} &=& -\sum_p \frac{\partial \epsilon_p^\text{HF}}{\partial \mu^\text{QP(2)}} f_p^- + \sum_p f_p^- 
\nonumber\\&&
 - \sum_p \left( \epsilon_p^\text{HF} - \mu^\text{QP(2)} \right) \frac{\partial f_p^-}{\partial \mu^\text{QP(2)}}
\nonumber\\&&
+ \sum_{p,q} \langle pq || pq  \rangle \frac{\partial f_p^-}{\partial \mu^{\text{QP(2)}}} f_q^-
\nonumber\\&&
 -\frac{1}{\beta} \sum_p \frac{\partial f_p^-}{\partial \mu^\text{QP(2)}} \ln f_p^- -\frac{1}{\beta} \sum_p \frac{\partial f_p^-}{\partial \mu^\text{QP(2)}}
 \nonumber\\&&
-\frac{1}{\beta} \sum_p \frac{\partial f_p^+}{\partial \mu^\text{QP(2)}} \ln f_p^+ -\frac{1}{\beta} \sum_p \frac{\partial f_p^+}{\partial \mu^\text{QP(2)}}  - \frac{\partial \langle E^{(2)} \rangle}{\partial \mu^\text{QP(2)}}
\nonumber\\&=&
\sum_p f_p^-  - \sum_p \left( \epsilon_p^\text{HF} - \mu^\text{QP(2)} \right) \frac{\partial f_p^-}{\partial \mu^\text{QP(2)}}
\nonumber\\&&
 -\frac{1}{\beta} \sum_p \frac{\partial f_p^-}{\partial \mu^\text{QP(2)}}  \ln \frac{f_p^-}{f_p^+} - \frac{\partial \langle E^{(2)} \rangle}{\partial \mu^\text{QP(2)}}
\nonumber\\&=&
\sum_p f_p^-  + \sum_p \Sigma_{pp}^{(2)}  \frac{\partial f_p^-}{\partial \mu^\text{QP(2)}} - \frac{\partial \langle E^{(2)}\rangle}{\partial \mu^\text{QP(2)}}
 \nonumber\\&=& \sum_p f_p^- 
 = \bar{N}, \label{QP:relation1explicit}
\end{eqnarray}
where we used
\begin{eqnarray}
\frac{\partial f_p^+}{\partial \mu^\text{QP(2)}} &=& - \frac{\partial f_p^-}{\partial \mu^\text{QP(2)}} , \label{minusplus} \\
\ln \frac{f_p^-}{f_p^+} &=& -\beta \left( \epsilon_p^\text{QP(2)} - \mu^\text{QP(2)} \right), \label{lnfminusfplus}
\end{eqnarray}
but not the explicit derivative $\partial f_p^- / \partial \mu^\text{QP(2)}$, which is
a solution of a system of linear equations and cannot be written in a closed form.\cite{Gu24}
This justifies the condition for $\mu^\text{QP(2)}$ [Eq.\ (\ref{QP:mu})].

In the penultimate equality of Eq.\ (\ref{QP:relation1explicit}), we also used
\begin{eqnarray}
\sum_p  \Sigma_{pp}^{(2)}\frac{\partial f_p^-} {\partial \mu^\text{QP(2)}}  = \frac{\partial \langle E^{(2)} \rangle}{\partial \mu^\text{QP(2)}}, \label{condition1}
\end{eqnarray}
which can be verified by using Eqs.\ (\ref{E2}) and (\ref{QP:sigma2_e0}).
In fact, the form of $\Sigma_{pp}^{(2)}$ is determined such that the above equation is satisfied in the first place. Generally,
\begin{eqnarray}
 \Sigma_{pp}^{(n)} \equiv \frac{\partial \langle E^{(n)} \rangle}{\partial f_p^- }, \label{selfenergy}
\end{eqnarray}
which ensures that Eq.\ (\ref{condition1}) and all similar relations hold. 

Equation (\ref{selfenergy}) is a finite-temperature generalization\cite{matsubara,luttingerward} of the $n$th-order self-energy of Feynman--Dyson 
MBGF in the diagonal, frequency-independent approximation.\cite{Hirata2017,deltamp} 
Owing to this diagonal construction, the second- and higher-order thermal quasi-particle theories do not involve 
rotation of orbitals, unlike thermal HF theory\cite{Mermin63,Gu24} or Feynman--Dyson MBGF without the diagonal approximation,\cite{Hirata2017} 
which define the HF or Dyson orbitals\cite{OrtizDyson} 
as those that bring the Fock or self-energy matrix into a diagonal form. Despite the absence of orbital rotation, $\Omega^\text{QP(2)}$ is still minimized by varying $f_p^-$ (see below).  

Diagrammatically, the differentiation with respect to $f_p^-$ corresponds to opening a closed, internal-energy diagram
by deleting an edge (whose mathematical interpretation is $f_p^\pm$).\cite{Hirata2017} 
A physical meaning of thermal self-energies or thermal quasi-particle energies, which are temperature dependent, is discussed in Sec.\ \ref{sec:thermalQPenergies}.

The following thermodynamic relation,
\begin{eqnarray}
 -\frac{\partial \Omega^\text{QP(2)}}{\partial T} &=& S^\text{QP(2)} , \label{QP:relation2}
\end{eqnarray}
is also obeyed. This too can be confirmed by explicit differentiation.
\begin{eqnarray}
-\frac{\partial \Omega^\text{QP(2)}}{\partial T} &=& k_{\text{B}}\beta^2 \frac{\partial\Omega^\text{QP(2)}}{\partial \beta} 
\nonumber\\
&=& k_{\text{B}}\beta^2 \sum_p  \frac{\partial  \epsilon_p^\text{HF} }{\partial \beta} f_p^- 
\nonumber\\&&
+ k_{\text{B}}\beta^2 \sum_p \left(\epsilon_p^\text{HF} - \mu^\text{QP(2)}\right) \frac{\partial f_p^-}{\partial \beta} 
\nonumber\\&& 
- k_{\text{B}}\beta^2 \sum_{p,q} \langle pq || pq  \rangle \frac{\partial f_p^-}{\partial \beta} f_q^-
\nonumber\\&& 
- k_{\text{B}}  \left( f_p^- \ln f_p^- + f_p^+ \ln f_p^+  \right)
\nonumber\\
&& + k_{\text{B}}\beta \sum_p \frac{\partial f_p^-}{\partial \beta} \ln \frac{f_p^-}{f_p^+}  
\nonumber\\ &&
+ k_{\text{B}}\beta^2 \frac{\partial \langle E^{(2)} \rangle}{\partial \beta} \nonumber \\
&=& - k_{\text{B}}  \left( f_p^- \ln f_p^- + f_p^+ \ln f_p^+  \right)
\nonumber\\ &&
- k_{\text{B}}\beta^2 \sum_p \Sigma_{pp}^{(2)} \frac{\partial f_p^-  }{\partial \beta} 
+ k_{\text{B}}\beta^2 \frac{\partial \langle E^{(2)}\rangle}{\partial \beta} \nonumber \\
&=& - k_{\text{B}}  \left( f_p^- \ln f_p^- + f_p^+ \ln f_p^+  \right) = S^\text{QP(2)}, \label{FD:relation2explicit}
\end{eqnarray}
where we used Eqs.\ (\ref{minusplus}), (\ref{lnfminusfplus}), and (\ref{selfenergy}), but not the explicit derivative $\partial f_p^- / \partial \beta$.\cite{Gu24}
This justifies the entropy formula [Eq.\ (\ref{QP:S})].

It can also be shown that thermal QP(2) theory is variational with respect to the diagonal elements of its one-electron density matrix, 
but not with respect to orbitals.
\begin{eqnarray}
\frac{\partial \Omega^\text{QP(2)}}{\partial f_p^-} &=& 
\left( \epsilon_p^\text{HF} -\mu^\text{QP(2)} \right) + \sum_q \frac{\partial \epsilon_q^\text{HF}}{\partial f_p^-} f_q^- 
- \sum_q \langle pq|| pq\rangle f_q^- \nonumber\\
&& + \frac{1}{\beta} \ln \frac{f_p^-}{f_p^+} + \frac{\partial \langle E^{(2)} \rangle }{\partial f_p^-}
\nonumber\\
&=& \left( \epsilon_p^\text{HF} -\mu^\text{QP(2)} \right) - \left( \epsilon_p^\text{QP(2)} -\mu^\text{QP(2)} \right) + \Sigma_{pp}^{(2)} = 0, \nonumber\\ \label{QP:variational}
\end{eqnarray}
where Eq.\ (\ref{selfenergy}) was used. This proves that the $f_p^-$ functions that minimize $\Omega^\text{QP(2)}$ take the forms of the Fermi--Dirac distribution functions, 
justifying Eqs.\ (\ref{QP:minus})--(\ref{QP:epsilon}). 

Insofar as Eq.\ (\ref{selfenergy}) is satisfied (and thus all thermodynamic relations are obeyed and the variationality is ensured), there is considerable latitude in selecting the functional
form of an approximate internal energy. For instance, the most notable departure of $U^\text{QP(2)}$ in Eq.\ (\ref{QP:U}) from the lengthy, but full $U^{(2)}$  of Eq.\ (74) of Ref.\ \onlinecite{HirataJhaJCP2020} 
is the absence of the so-called anomalous-diagram terms,\cite{kohn,luttingerward} 
which sum over the set of indexes whose fictitious denominator is zero.\cite{HirataJhaJCP2020,Hirata2021}  
They are purposefully neglected in our ansatz because they are found to cause a severe Kohn--Luttinger nonconvergence problem\cite{kohn,luttingerward,Hirata_KL2021,Hirata_KL2022}
in $\Sigma_{pp}^{(2)}$, and are undesirable. This issue will be expounded in Appendix \ref{sec:appendix}.

When the reference wave function is supplied by a finite-temperature theory, its $\epsilon_p^{(0)}$ depends on temperature and the formulas for the thermal self-energies need to be adjusted accordingly. In the thermal HF reference, the {\it second-order} thermal self-energy $\Sigma_{pp}^{(2)}$ is found to contain {\it third-order} energy-independent terms.\cite{purvis_cpl,SchirmerAngonoa}  
Although they seem no more expensive to evaluate than the rest, we will not consider this reference any further in this initial study.

Not only is finite-temperature MBPT plagued by the Kohn--Luttinger nonconvergence,\cite{Hirata_KL2021,Hirata_KL2022}  
but the Feynman--Dyson perturbation series of zero-temperature MBGF is also shown to display even severer divergence for most low- and high-lying states.\cite{Hirata_PRA2024} 
Potential impact of the latter divergence on thermal quasi-particle theories will be contemplated in Appendix \ref{sec:appendix2}. 


\section{Numerical results}
\subsection{Thermodynamic functions}

\begin{table}
\caption{\label{table:Omega} Grand potential $\Omega$ (in $E_\text{h}$) of an ideal gas of the identical hydrogen fluoride molecules (0.9168\,\AA) in the STO-3G basis set as a function of temperature ($T$).}
\begin{ruledtabular}
\begin{tabular}{crrrr}
$T$ / K & \multicolumn{1}{c}{HF\tablenotemark[1]} & \multicolumn{1}{c}{MBPT(2)\tablenotemark[2]} & \multicolumn{1}{c}{QP(2)\tablenotemark[3]} & \multicolumn{1}{c}{FCI\tablenotemark[4]} \\ \hline
$10^4$ &   $-99.50758$ &   $-99.94001$ &   $-99.94179$ &   $-99.94377$ \\
$10^5$ &  $-101.02137$ &  $-103.48646$ &  $-101.30202$ &  $-102.10659$ \\
$10^6$ &  $-150.56294$ &  $-151.43748$ &  $-150.60284$ &  $-151.24440$ \\
$10^7$ &  $-729.93806$ &  $-730.10421$ &  $-729.94666$ &  $-730.09519$ \\
$10^8$ & $-6846.98049$ & $-6847.00261$ & $-6846.98165$ & $-6847.00247$ \\
\end{tabular}
\tablenotetext[1]{Thermal HF theory.}
\tablenotetext[2]{Second-order finite-temperature MBPT\cite{HirataJha,HirataJhaJCP2020,Hirata2021} with zero-temperature HF reference.}
\tablenotetext[3]{Second-order thermal quasi-particle theory with zero-temperature HF reference.}
\tablenotetext[4]{Thermal FCI theory.\cite{Kou}}
\end{ruledtabular}
\end{table}

\begin{figure}
  \includegraphics[width=\columnwidth]{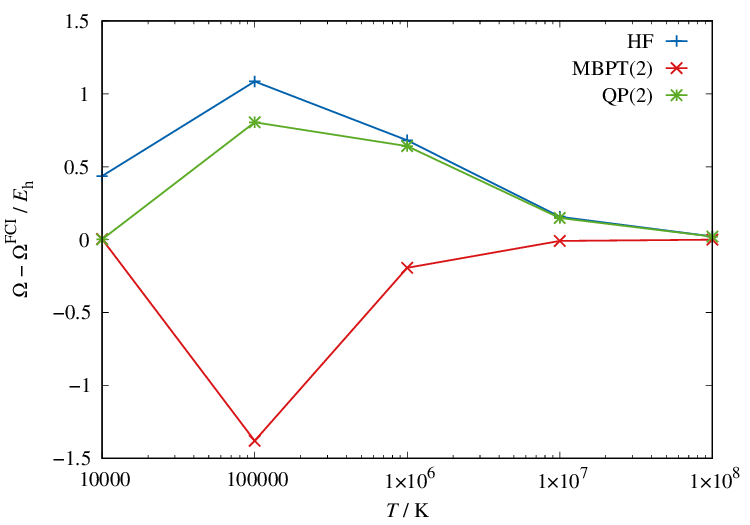}
\caption{The deviation from thermal FCI theory in grand potential $\Omega$ of an ideal gas of the identical hydrogen fluoride molecules (0.9168\,\AA) in the STO-3G basis set as a function of temperature ($T$). The ``HF'' stands 
for thermal HF theory, ``MBPT(2)'' denotes second-order finite-temperature many-body perturbation theory with the zero-temperature HF reference, ``QP(2)'' 
designates second-order thermal quasi-particle theory with the zero-temperature HF reference. }
\label{fig:FH_Omega}
\end{figure}

\begin{table}
\caption{\label{table:U} Same as Table \ref{table:Omega} but for internal energy $U$ (in $E_\text{h}$).}
\begin{ruledtabular}
\begin{tabular}{crrrr}
$T$ / K & \multicolumn{1}{c}{HF\tablenotemark[1]} & \multicolumn{1}{c}{MBPT(2)\tablenotemark[2]} & \multicolumn{1}{c}{QP(2)\tablenotemark[3]} & \multicolumn{1}{c}{FCI\tablenotemark[4]} \\ \hline
$0$\tablenotemark[5] & $-98.57076$ & $-98.58809$ & $-98.58809$ & $-98.59658$ \\
$10^4$ & $-98.57076$ & $-98.58809$ & $-98.58809$ & $-98.59658$ \\
$10^5$ & $-97.94385$ & $-97.86604$ & $-97.97596$ & $-98.04938$ \\
$10^6$ & $-96.79410$ & $-96.99284$ & $-96.80270$ & $-96.94534$ \\
$10^7$ & $-92.02773$ & $-92.05724$ & $-92.02910$ & $-92.05557$ \\
$10^8$ & $-88.48266$ & $-88.48744$ & $-88.48288$ & $-88.48740$ \\
\end{tabular}
\tablenotetext[1]{$^{-\text{ d}}$ See the corresponding footnotes of Table \ref{table:Omega}.}
\tablenotetext[5]{Ground-state energy from zero-temperature HF, MBPT(2), and FCI theories.}
\end{ruledtabular}
\end{table}

\begin{figure}
  \includegraphics[width=\columnwidth]{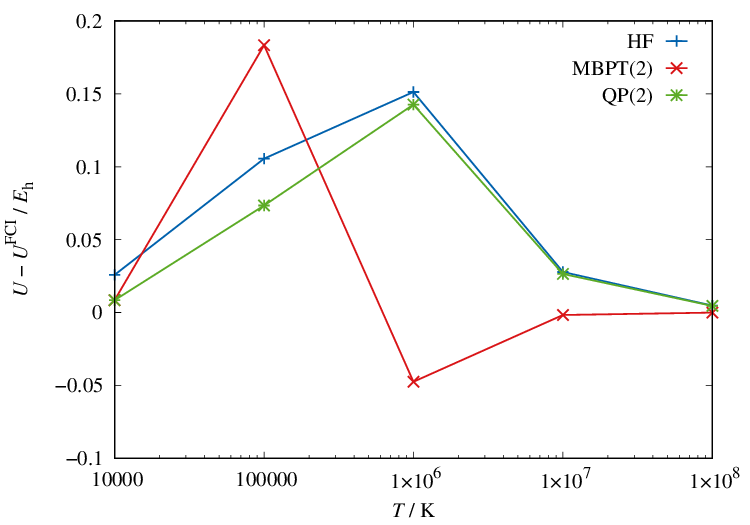}
\caption{Same as Fig.\ \ref{fig:FH_Omega} but for internal energy $U$.}
\label{fig:FH_U}
\end{figure}

\begin{table}
\caption{\label{table:mu} Same as Table \ref{table:Omega} but for chemical potential $\mu$ (in $E_\text{h}$).}
\begin{ruledtabular}
\begin{tabular}{crrrr}
$T$ / K & \multicolumn{1}{c}{HF\tablenotemark[1]} & \multicolumn{1}{c}{MBPT(2)\tablenotemark[2]} & \multicolumn{1}{c}{QP(2)\tablenotemark[3]} & \multicolumn{1}{c}{FCI\tablenotemark[4]} \\ \hline
$10^4$ &   $0.09368$ &   $0.13519$ &   $0.13537$ &   $0.13472$ \\
$10^5$ &   $0.20722$ &   $0.42903$ &   $0.23246$ &   $0.29568$ \\
$10^6$ &   $3.80022$ &   $3.87744$ &   $3.80378$ &   $3.85990$ \\
$10^7$ &  $46.85490$ &  $46.86975$ &  $46.85568$ &  $46.86892$ \\
$10^8$ & $504.65280$ & $504.65478$ & $504.65291$ & $504.65476$ \\
\end{tabular}
\tablenotetext[1]{$^{-\text{ d}}$ See the corresponding footnotes of Table \ref{table:Omega}.}
\end{ruledtabular}
\end{table}

\begin{figure}
  \includegraphics[width=\columnwidth]{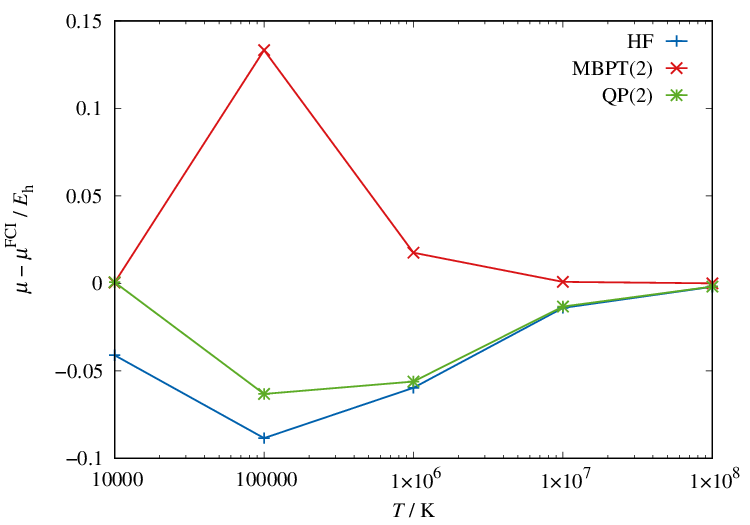}
\caption{Same as Fig.\ \ref{fig:FH_Omega} but for chemical potential $\mu$.}
\label{fig:FH_mu}
\end{figure}

\begin{table}
\caption{\label{table:S} Same as Table \ref{table:Omega} but for entropy $S$ (in $k_\text{B}$).}
\begin{ruledtabular}
\begin{tabular}{crrrr}
$T$ / K & \multicolumn{1}{c}{HF\tablenotemark[1]} & \multicolumn{1}{c}{MBPT(2)\tablenotemark[2]} & \multicolumn{1}{c}{QP(2)\tablenotemark[3]} & \multicolumn{1}{c}{FCI\tablenotemark[4]} \\ \hline
$10^4$ & $0.00000$ & $0.00001$ & $0.00001$ & $0.00011$ \\
$10^5$ & $3.17451$ & $4.20017$ & $3.16235$ & $3.47472$ \\
$10^6$ & $4.97871$ & $4.94828$ & $4.97736$ & $4.95769$ \\
$10^7$ & $5.34800$ & $5.34763$ & $5.34798$ & $5.34766$ \\
$10^8$ & $5.40597$ & $5.40596$ & $5.40597$ & $5.40596$ \\
\end{tabular}
\tablenotetext[1]{$^{-\text{ d}}$ See the corresponding footnotes of Table \ref{table:Omega}.}
\end{ruledtabular}
\end{table}

\begin{figure}
  \includegraphics[width=\columnwidth]{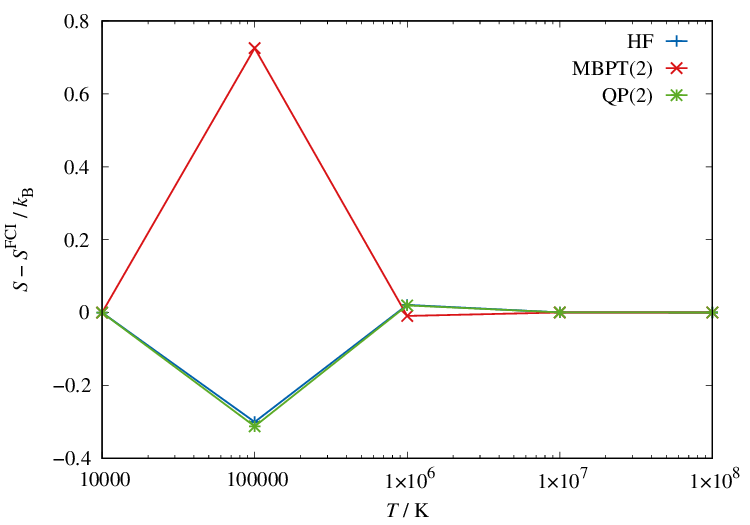}
\caption{Same as Fig.\ \ref{fig:FH_Omega} but for entropy $S$.}
\label{fig:FH_S}
\end{figure}

Tables \ref{table:Omega}--{\ref{table:S} list the thermodynamic functions---grand potential $\Omega$, internal energy $U$, chemical potential $\mu$, and entropy $S$---of 
thermal HF [thermal QP(1)],\cite{Gu24} thermal QP(2), fintie-temperature MBPT(2),\cite{HirataJhaJCP2020,Hirata2021} and thermal full-configuration-interaction (FCI) theories\cite{Kou} for 
an ideal gas of the identical hydrogen fluoride molecules in a wide range of temperature ($T$). Figures \ref{fig:FH_Omega}--\ref{fig:FH_S} plot the deviations from thermal FCI benchmarks, the latter being 
exact within a basis set.
The thermal QP(2) and finite-temperature MBPT(2) calculations were based on the zero-temperature HF reference.
See Ref.\ \onlinecite{Gu24} for the data for the same system from Fermi--Dirac [thermal QP(0)] theory and other thermal mean-field theories, which do not take into account electron correlation.

In all cases, the thermal QP(2) results are close to finite-temperature MBPT(2) ones at low $T$ and are more accurate than thermal HF theory. 
This is expected because both thermal QP(2) theory and finite-temperature MBPT(2) reduce to zero-temperature MBPT(2) for internal energy, accounting for electron correlation, while thermal and zero-temperature HF theories do not. At higher $T$, thermal QP(2) theory quickly approaches thermal HF theory, where their quasi-independent-particle picture becomes relatively less erroneous. However,
at $T=10^6$\,K and $10^7$\,K, finite-temperature MBPT(2) is systematically more accurate as it accounts for the effect of electron correlation in $S$ and $\mu$ and it uses more complete second-order corrections 
(including anomalous-diagram terms) in $\Omega$ and $U$. At the intermediate temperature of $T=10^5$\,K, thermal QP(2) theory (or even thermal HF theory) outperforms finite-temperature MBPT(2).
This may be ascribed to the fact that thermal QP(2) theory includes the effect of electron correlation in the orbital energies, and thus higher-order correlation corrections in a manner similar in spirit to
self-consistent Green's function methods,\cite{luttingerward,BaymKadanoff1961,Baym_selfconsistent,VanNeck1991,Dickhoff_chapter7,Dickhoff2004,Dahlen2005,Barbieri2006,Barbieri2009,Phillips2014,Neuhauser2017,Tarantino2017,Kadanoff_book,CoveneyTew2023}
which are said to work for strong correlation. However, this similarity must not be overstated because self-consistent Green's function methods replace reference orbitals and orbital energies by electron-correlated counterparts, while thermal QP(2) theory uses correlated orbital energies only in the Fermi--Dirac distribution functions and thus reduces identically to MBPT(2) at $T=0$. 

Additional numerical data are presented in Supplementary Information,\cite{supp01} which reinforces the foregoing conclusions.

\subsection{Thermal quasi-particle energies\label{sec:thermalQPenergies}}

Equation (\ref{selfenergy}) can be rewritten as
\begin{eqnarray}
\epsilon_p^\text{QP} \equiv \frac{\partial U^\text{QP}}{\partial f_p^-}, \label{thermalJanak}
\end{eqnarray}
which encompasses all of the Fermi--Dirac, 
thermal HF, 
and thermal QP(2) ans\"{a}tze. 

Hence, the $\epsilon_p^\text{QP}$ has the literal physical meaning of the increase in
the internal energy upon infusion of an infinitesimal fraction of an electron in the $p$th spinorbital. This interpretation can be viewed as a finite-temperature version of 
Janak's theorem\cite{Janak1978} in DFT, and may be called thermal Janak's theorem. 
It is not unreasonable to consider a fraction of an electron here because the process in question is the thermal average
of the same processes involving an infinite number of particles.

While the $\epsilon_p^\text{QP}$ at $T = 0$ is rigorously related to the 
ionization and electron-attachment energies of a molecule or solid according to Koopmans' theorem\cite{szabo} or MBGF,\cite{Hirata2017} 
it cannot be expected to be even close to their thermal averages at $T \gg 0$ 
because the latter involve different orbitals at different probabilities, even in a strict independent-particle picture.  

Therefore, the simplest reasonable approximations to thermal ionization ($I^\text{QP}$) and electron-attachment ($A^\text{QP}$) energies (signs reversed) based on the $\epsilon_p^\text{QP}$ 
may be their weighted averages such as
\begin{eqnarray}
I^\text{QP} &=& \sum_p \epsilon_p^\text{QP} f_p^- (\bar N) - \sum_p \epsilon_p^\text{QP} f_p^- (\bar N-1) , \label{I}\\
A^\text{QP} &=& \sum_p \epsilon_p^\text{QP} f_p^- (\bar N+1) - \sum_p \epsilon_p^\text{QP} f_p^- (\bar N) , \label{A}
\end{eqnarray}
where $\bar N$ is the average number of electrons that ensures electroneutrality and $f_p^- (N)$ is the Fermi--Dirac distribution function for an arbitrary (integer or noninteger) average number of electrons $N$, i.e.,
\begin{eqnarray}
N = \sum_p f_p^- (N).
\end{eqnarray}
It should be understood that only the chemical potential $\mu^\text{QP}$ (but not $\epsilon_p^\text{QP}$) is varied so that this equation is satisfied. 

We examine the validity of thermal Janak's theorem by comparing the above approximations ($I^\text{QP}$ and $A^\text{QP}$) against $\Delta U^\text{FCI}$ defined by 
\begin{eqnarray}
\Delta U^\text{FCI} &\equiv& \left\{ \begin{array}{ll} U^\text{FCI}(\bar N)-U^\text{FCI}(\bar N-1), & \text{ionization}; \\ U^\text{FCI}(\bar N+1)-U^\text{FCI}(\bar N), & \text{electron attachment}, \end{array} \right. \nonumber\\
\end{eqnarray}
where 
$U^\text{FCI}(N)$ is the internal energy of thermal FCI theory for an ideal gas of identical molecules, each of which has $N$ electrons on average. 
It is 
\begin{eqnarray}
U^\text{FCI}(N) &\equiv& \frac{\sum_I E_I e^{-\beta(E_I - \mu N_I) } } {\sum_I  e^{-\beta(E_I - \mu N_I) } },
\end{eqnarray}
where $E_I$ is the FCI energy of the $I$th state and the chemical potential $\mu$ is determined by
\begin{eqnarray}
N &=&  \frac{\sum_I N_I e^{-\beta(E_I - \mu N_I) } } {\sum_I  e^{-\beta(E_I - \mu N_I) } }. \label{N}
\end{eqnarray}
This is a valid and in fact, exact (within a basis set) treatment of thermodynamics if the particles are not electrically charged. 
For electrons, however, it describes a physically unrealistic, massively charged plasma, whose energy is not even extensive.\cite{Fisher,Dyson,HirataOhnishi}
Nevertheless, $U^\text{FCI}(N)$ and $\Delta U^\text{FCI}$ are computationally well defined even for electrons thanks to the ideal-gas assumption (i.e.,
no interactions between molecules) and will be useful for the following analysis. It may or may not be a reasonable approximation to the thermal average of ionization or electron-attachment energies.

The approximations of Eqs.\ (\ref{I}) and (\ref{A}) 
imply the corresponding internal-energy functional of the form,
\begin{eqnarray}
U^\text{QP}(N) \equiv \sum_p \epsilon_p^\text{QP} f_p^-(N),
\end{eqnarray}
which leads to
\begin{eqnarray}
\Delta U^\text{QP} &\equiv& \left\{ \begin{array}{ll} I^\text{QP} , & \text{ionization}; \\ A^\text{QP}, & \text{electron attachment}. \end{array} \right. 
\end{eqnarray}

The $U^\text{FCI}(N)$ and $U^\text{QP}(N)$ thus defined are computed for an ideal gas of the identical hydrogen fluoride molecules by thermal FCI, thermal QP(2), and thermal HF theories and are plotted as a function of $N$
in Figs.\ \ref{fig:FH_FCI_U}, \ref{fig:FH_QP2_U}, and \ref{fig:FH_HF_U}, respectively. The thermal ionization energies of these three methods 
are compared with one another and also with the thermal HOMO energies in Table \ref{table:IP}. 
The same data for electron-attachment energies and LUMO are compiled in Table \ref{table:EA}. 

\begin{figure}
\includegraphics[width=\columnwidth]{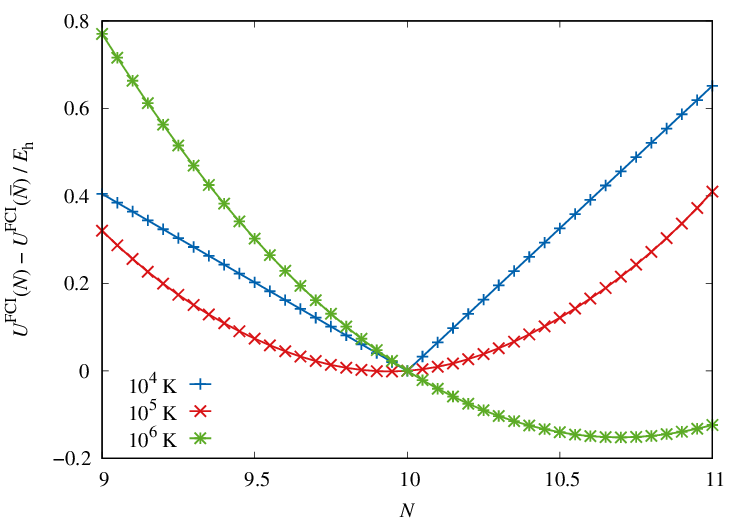}
\caption{Internal energy of thermal FCI theory $U^\text{FCI}$ (in $E_\text{h}$) of an ideal gas of the identical hydrogen fluoride molecules (0.9168\,\AA) in the STO-3G basis set  as a function of the average number of electrons $N$ relative to the value at $N=\bar{N}=10$. }
\label{fig:FH_FCI_U}
\end{figure}

\begin{figure}
\includegraphics[width=\columnwidth]{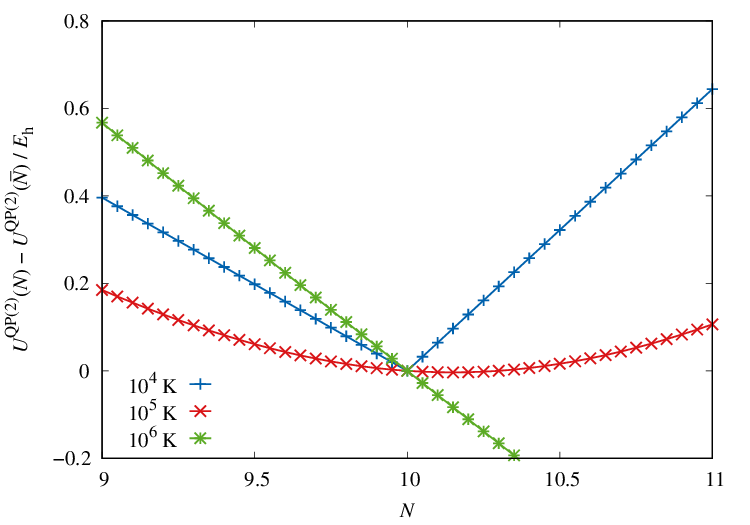}
\caption{Same as Fig.\ \ref{fig:FH_FCI_U} but for the internal energy  of thermal QP(2) theory $U^\text{QP(2)}$.}
\label{fig:FH_QP2_U}
\end{figure}

\begin{figure}
\includegraphics[width=\columnwidth]{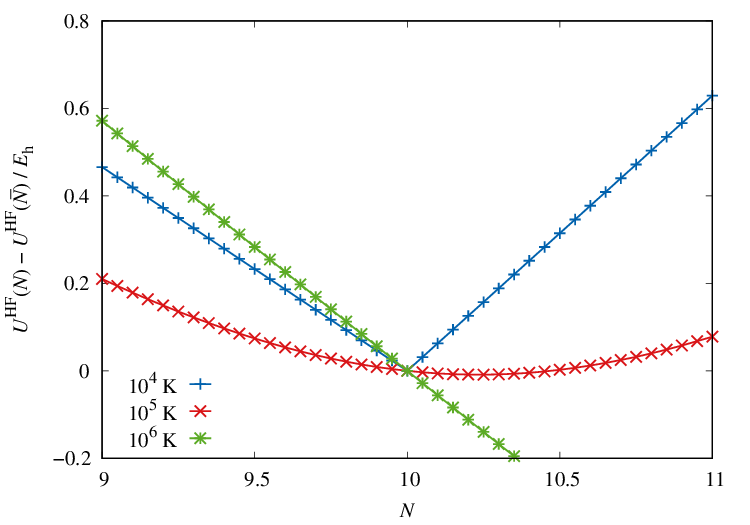}
\caption{Same as Fig.\ \ref{fig:FH_FCI_U} but for the internal energy of thermal HF theory $U^\text{HF}$.}
\label{fig:FH_HF_U}
\end{figure}

\begin{table}
\caption{\label{table:IP} Thermal ionization energy (in $E_\text{h}$) of an ideal gas of the identical hydrogen fluoride molecules (0.9168\,\AA) in the STO-3G basis set as a function of temperature ($T$).}
\begin{ruledtabular}
\begin{tabular}{crrrrr}
 & \multicolumn{2}{c}{$\epsilon_p^\text{QP}$} & \multicolumn{3}{c}{$\Delta U$} \\ \cline{2-3}\cline{4-6}
$T$ / K & \multicolumn{1}{c}{HF\tablenotemark[1]} & \multicolumn{1}{c}{QP(2)\tablenotemark[2]} & \multicolumn{1}{c}{HF\tablenotemark[1]} & \multicolumn{1}{c}{QP(2)\tablenotemark[2]} & \multicolumn{1}{c}{FCI\tablenotemark[3]}  \\ \hline
$0$\tablenotemark[4]     & $-0.46417$ & $-0.39557$ & $-0.46417$ & $-0.39557$ & $-0.40429$ \\
$10^4$  & $-0.46417$ & $-0.39557$ & $-0.46589$ & $-0.39603$ & $-0.40468$ \\
$10^5$  & $-0.45147$ & $-0.41998$ & $-0.21004$ & $-0.18483$ & $-0.32041$ \\
$10^6$  & $-0.57384$ & $-0.57392$ & $-0.57181$ & $-0.56735$ & $-0.77028$ \\
$10^7$  & $-0.69361$ & $-0.69551$ & $-3.40146$ & $-3.40055$ & $-3.65153$ \\
$10^8$  & $-0.76988$ & $-0.77193$ & $-4.95141$ & $-4.95127$ & $-5.34456$ \\
\end{tabular}
\tablenotetext[1]{Thermal HF theory.}
\tablenotetext[2]{Second-order thermal quasi-particle theory with zero-temperature HF reference.}
\tablenotetext[3]{Thermal FCI theory.\cite{Kou}}
\tablenotetext[4]{Ionization energy from the corresponding MBGF.\cite{Hirata2017,Hirata_PRA2024}}
\end{ruledtabular}
\end{table}

\begin{table}
\caption{\label{table:EA} Thermal electron-attachment energy (in $E_\text{h}$) of an ideal gas of the identical hydrogen fluoride molecules (0.9168\,\AA) in the STO-3G basis set as a function of temperature ($T$).}
\begin{ruledtabular}
\begin{tabular}{crrrrr}
 & \multicolumn{2}{c}{$\epsilon_p^\text{QP}$} & \multicolumn{3}{c}{$\Delta U$} \\ \cline{2-3}\cline{4-6}
$T$ / K & \multicolumn{1}{c}{HF\tablenotemark[1]} & \multicolumn{1}{c}{QP(2)\tablenotemark[2]} & \multicolumn{1}{c}{HF\tablenotemark[1]} & \multicolumn{1}{c}{QP(2)\tablenotemark[2]} & \multicolumn{1}{c}{FCI\tablenotemark[3]}  \\ \hline
$0$\tablenotemark[4]      & $0.62924$ & $0.64424$ &  $0.62924$ &  $0.64424$ &  $0.65170$ \\
$10^4$  & $0.62924$ & $0.64424$ &  $0.62924$ &  $0.64424$ &  $0.65170$ \\
$10^5$  & $0.48080$ & $0.50816$ &  $0.07823$ &  $0.10621$ &  $0.40988$ \\
$10^6$  & $0.28118$ & $0.31458$ & $-0.55009$ & $-0.54444$ & $-0.12383$ \\
$10^7$  & $0.23384$ & $0.27782$ & $-3.14063$ & $-3.13959$ & $-2.71365$ \\
$10^8$  & $0.21118$ & $0.26168$ & $-4.90424$ & $-4.90408$ & $-4.48208$ \\
\end{tabular}
\tablenotetext[1]{Thermal  HF theory.}
\tablenotetext[2]{Second-order thermal quasi-particle theory with zero-temperature HF reference.}
\tablenotetext[3]{Thermal FCI theory.\cite{Kou}}
\tablenotetext[4]{Electron-attachment energy from the corresponding MBGF.\cite{Hirata2017,Hirata_PRA2024}}
\end{ruledtabular}
\end{table}

From the tables, we observe that while the HOMO and LUMO energies of thermal quasi-particle theories correctly reduce to the ionization and electron-attachment 
energies of MBGF in the $T=0$ limit, they clearly have nothing to do with the approximate thermal ionization or electron-attachment energies  
as defined by $\Delta U^\text{FCI}$
at $T \gg 0$, notwithstanding the questionable validity of the latter. However, $\Delta U^\text{QP}$ calculated with $\epsilon_p^\text{QP}$ 
are in line with the approximate thermal ionization or electron-attachment energies, giving credence to the physical meaning
of thermal orbital energies provided by thermal Janak's theorem. 

Thermal QP(2) theory 
is much closer to thermal FCI theory than thermal HF theory at low temperatures ($T \leq 10^4$\,K), correctly accounting for electron correlation in thermal ionization and electron-attachment energies.
However, at higher $T$, thermal HF and QP(2) theories are essentially the same, 
suggesting that the effect of electron correlation is surpassed by the lack thereof in the entropy and chemical-potential terms that dominate at such temperatures.

The figures reinforce these observations, but provide the following additional insights: As $T$ is lowered, all $U(N)$ curves begin to display signs of a derivative discontinuity at $N=\bar{N}$, although
the strict discontinuity does not occur until $T=0$. Recall that all nonhybrid DFT approximations lack a derivative discontinuity,\cite{perdew_1983} causing a gross underestimation of band gaps.
In both the low- and high-$T$ extremes, $U(N)$ has near-linear dependence on $N$ within each interval between adjacent integers, but at intermediate $T$, 
each $U(N)$ curve is convex  with its minimum occurring away from $N=\bar{N}$. 

Let us consider the slope of $U^\text{QP}(N)$ at $N=\bar{N}$, which is another macroscopic quantity evaluable with $\epsilon_p^\text{QP}$. 
\begin{eqnarray}
\frac{\partial U^\text{QP}}{\partial  N} &=&  \frac{\partial \mu^\text{QP}}{\partial  N}  \sum_p \frac{\partial f_p^-}{\partial \mu^\text{QP}}\frac{\partial U^\text{QP}}{\partial f_p^-} \nonumber\\
&=& \frac{\sum_p f_p^- f_p^+ \epsilon_p^\text{QP}}{\sum_p f_p^- f_p^+ }, \label{dUdN}
\end{eqnarray}
where we used Eq.\ (\ref{N}) and 
\begin{eqnarray}
\frac{\partial f_p^-}{\partial \mu^\text{QP}} = \beta f_p^- f_p^+.
\end{eqnarray}
This is the increase in the internal energy upon infusion of an infinitesimal fraction of an electron into every molecule in the ideal gas.
It is a weighted average of $\epsilon_p^\text{QP}$ over all spinorbitals. It corresponds to neither  the ionization nor  electron-attachment process
because it is oblivious to the sign of the variation in $N$. However, in the $T=0$ limit, it distinguishes these two processes by exhibiting the derivative discontinuity, i.e.,
\begin{eqnarray}
\lim_{T \to 0} \left( \frac{\partial U^\text{QP}}{\partial  N} \right)_{N=\bar{N}^-} &=& \epsilon_\text{HOMO}^\text{QP}, \\
\lim_{T \to 0} \left( \frac{\partial U^\text{QP}}{\partial  N} \right)_{N=\bar{N}^+} &=& \epsilon_\text{LUMO}^\text{QP}. 
\end{eqnarray}
Until the $T=0$ limit is strictly reached, however, the $U^\text{QP}(N)$ curve is smooth everywhere and the slope 
approaches the midpoint of the HOMO and LUMO energies.\cite{perdew_1983} For $T\approx 0$,
\begin{eqnarray}
 \left( \frac{\partial U^\text{QP}}{\partial  N} \right)_{N=\bar{N}} &\approx& \frac{ \epsilon_\text{HOMO}^\text{QP} + \epsilon_\text{LUMO}^\text{QP} }{2}, 
\end{eqnarray}
which is also equal to the $T=0$ limit of the $\mu^\text{QP}$.\cite{HirataJha,Hirata_KL2021}

\begin{table}
\caption{\label{table:IPEA} $\partial U / \partial N$ (in $E_\text{h}$) of an ideal gas of the identical hydrogen fluoride molecules (0.9168\,\AA) in the STO-3G basis set as a function of temperature ($T$). }
\begin{ruledtabular}
\begin{tabular}{crrr}
$T$ / K & \multicolumn{1}{c}{HF\tablenotemark[1]} & \multicolumn{1}{c}{QP(2)\tablenotemark[2]} & \multicolumn{1}{c}{FCI\tablenotemark[3]}  \\ \hline
$0$\tablenotemark[4]     &  $0.08253$ &  $0.12433$ &  $0.12371$ \\
$10^4$ &  $0.08189$ &  $0.12461$ &  $0.12351$ \\
$10^5$ & $-0.07423$ & $-0.04741$ &  $0.04959$ \\
$10^6$ & $-0.56092$ & $-0.55587$ & $-0.44097$ \\
$10^7$ & $-3.26523$ & $-3.26425$ & $-3.17327$ \\
$10^8$ & $-4.92771$ & $-4.92757$ & $-4.91206$ \\
\end{tabular}
\tablenotetext[1]{Thermal HF theory [Eq.\ (\ref{dUdN})].}
\tablenotetext[2]{Second-order thermal quasi-particle theory with zero-temperature HF reference [Eq.\ (\ref{dUdN})].}
\tablenotetext[3]{Thermal FCI theory (numerical differentiation).\cite{Kou}}
\tablenotetext[4]{The average of the ionization and electron-attachment energies (signs reversed) of the corresponding MBGF.\cite{Hirata2017,Hirata_PRA2024}}
\end{ruledtabular}
\end{table}

Table \ref{table:IPEA} compares the slope $\partial U/\partial N$ at $N=\bar{N}$ computed analytically by Eq.\ (\ref{dUdN}) for 
thermal HF and QP(2) theories against the slope of thermal FCI theory obtained by numerical differentiation. 
The three sets of values are in reasonable agreement with one another at all $T$ considered. At low $T$, thermal QP(2) theory
is in much better agreement with thermal FCI theory than thermal HF theory. This is because the former takes into account electron correlation. At higher $T$, 
thermal QP(2) and HF theories are more alike than thermal FCI theory, but they all seem to converge
at the same high $T$ limit. 

To summarize the results of the foregoing analysis, the orbital energies $\epsilon_p^\text{QP}$ of thermal quasi-particle theories (including thermal HF theory) cannot be directly related
to the ionization or electron-attachment energy from/into the $p$th orbital except at $T = 0$. This is simply because an ionization in an ideal gas of molecules at $T \gg 0$ 
is not a removal of an electron from the $p$th orbital in every molecule; rather, it is a removal of an electron from various orbitals of all molecules 
at some probabilities (even in a strict independent-particle picture). This is, in turn, equivalent to a removal of some fraction of an electron from various orbitals of an ``average'' molecule. 
The $\epsilon_p^\text{QP}$ at $T > 0$ signifies the increase in internal energy upon adding an infinitesimal fraction of an electron in the $p$th spinorbital of this average molecule. 
This interpretation may be called thermal Janak's theorem.\cite{Janak1978} 
This quantity may not directly correspond to any observable, but it is likely combined to approximate macroscopic thermodynamic observables, for which 
thermal QP(2) theory is expected to be more accurate than thermal HF theory.
A further analysis is needed to justify the use of these quasi-particle energies in characterizing, e.g., metal-insulator transitions, however.

\section{Conclusions}

The thermal quasi-particle theory has been introduced. It is a quasi-independent-particle theory and an electron-correlated extension 
of the widely used thermal HF theory. It is size-consistent and thus directly applicable to electron-correlated energy bands and thermodynamic functions of 
semiconductors and insulators at finite temperature.

Its entropy and chemical-potential formulas are of the one-electron type,
\begin{eqnarray}
S \equiv -k_\text{B} \sum_p \left( f_p^- \ln f_p^- + f_p^+ \ln f_p^+ \right), \nonumber
\end{eqnarray}
and
\begin{eqnarray}
\bar{N} \equiv \sum_p f_p^-, \nonumber
\end{eqnarray}
where the thermal population $f_p^-$ is also unchanged from the Fermi--Dirac distribution function,
\begin{eqnarray}
f_p^- \equiv \frac{1}{1+e^{\beta(\epsilon_p - \mu)}}, \nonumber
\end{eqnarray}
and $f_p^+ = 1-f_p^-$. The orbital energies $\epsilon_p$ now become the ones that include the effect of electron correlation by being 
defined as the correlated, thermal one-electron energies in the spirit of MBGF,
\begin{eqnarray}
\epsilon_p \equiv \frac{\partial U}{\partial f_p^-}, \nonumber
\end{eqnarray}
where $U$ is a correlated internal energy, whose mathematical form can be postulated on the basis of 
finite-temperature MBPT through any chosen order. When the grand potential is defined by
\begin{eqnarray}
\Omega \equiv U - \mu \bar{N} - T S, \nonumber
\end{eqnarray}
the theory satisfies all fundamental thermodynamic relations and the $\Omega$ is a variational minimum with respect to $f_p^-$. 

Fermi--Dirac and thermal HF theories can be viewed as the zeroth- and first-order instances, respectively, of the thermal quasi-particle theory hierarchy. 
It can be viewed as a diagonal, frequency-independent approximation to a finite-temperature extension of the inverse Dyson equation, which at zero temperature 
is an exact one-particle theory.\cite{Hirata2017} In other words, the $\epsilon_p$ thus defined is a finite-temperature extension of the Dyson self-energy in the diagonal, 
frequency-independent approximation. 

It has also been revealed that the thermal self-energy 
suffers from a severe Kohn--Luttinger nonconvergence problem, making it inappropriate to include anomalous-diagram terms 
in the definition of approximate $U$ or $\Omega$. This may mean that the thermal quasi-particle theory hierarchy is not convergent at exactness, unfortunately. 

Our preliminary implementation and comparison with the thermal FCI benchmarks suggests that thermal QP(2) theory 
performs distinctly better than thermal HF theory at low $T$ and may even outperform finite-temperature MBPT(2) at intermediate
$T$ by including electron-correlation effects in the orbital energies. At higher $T$, the entropy contribution dominates and the differences among 
various methods become relatively insignificant. The zero-temperature limit of thermal QP(2) theory is MBPT(2) for internal energy and MBGF(2) (in the diagonal, frequency-independent approximation) 
or $\Delta$MP2 for orbital energies.

The direct physical meaning of $\epsilon_p$ is the increase in the internal energy upon infusion of an infinitesimal fraction of an electron in the $p$th spinorbital, i.e., thermal Janak's theorem.\cite{Janak1978} 
It does not correspond to the ionization energy or electron-attachment energy of a gas of molecules or a solid at $T > 0$. However, it can be used as the key ingredients from which 
electron-correlated thermodynamic observables can be computed. 

\acknowledgments
This work was supported by the U.S. Department of Energy (DoE), Office of Science, Office of Basic Energy Sciences under Grant No.\ DE-SC0006028 and also by the Center for Scalable Predictive methods for Excitations and Correlated phenomena (SPEC), which is funded by the U.S. DoE, Office of Science, Office of Basic Energy Sciences, Division of Chemical Sciences, Geosciences and Biosciences as part of the Computational Chemical Sciences (CCS) program at Pacific Northwest National Laboratory (PNNL) under FWP 70942. PNNL is a multi-program national laboratory operated by Battelle Memorial Institute for the U.S. DoE.
The author is a Guggenheim Fellow of the John Simon Guggenheim Memorial Foundation. 

\appendix

\section{Kohn--Luttinger nonconvergence\label{sec:appendix}}

\begin{figure}
  \includegraphics[scale=0.4]{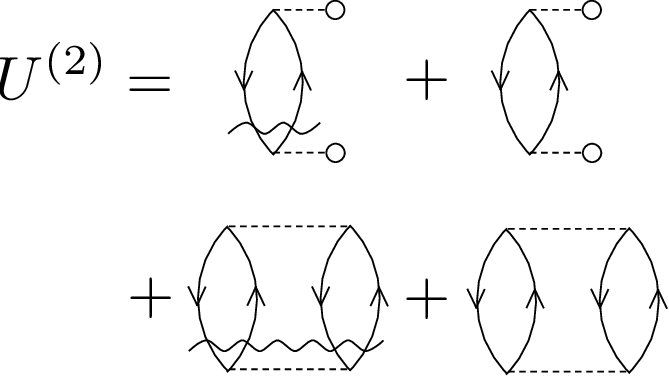}
\caption{Second-order internal energy diagrams of Eq.\ (\ref{app:U2}), which include anomalous (second and fourth) diagrams with no resolvent lines.}
\label{fig:U2}
\end{figure}

According to finite-temperature MBPT,\cite{HirataJha,HirataJhaJCP2020,Hirata2021} a more complete expression for the 
second-order internal energy should include the so-called anomalous-diagram contributions.\cite{kohn} It may read
\begin{eqnarray}
U^\text{QP} &\equiv& \langle E^{(0)} \rangle +  \langle E^{(1)} \rangle +  
 U^{(2)}  
\end{eqnarray}
with
\begin{eqnarray}
 U^{(2)}  &\equiv& \langle E^{(2)} \rangle - \beta \langle E^{(1)} E^{(1)} \rangle +  \beta \langle E^{(1)} \rangle\langle E^{(1)} \rangle \nonumber\\
 &=& \langle E^{(2)} \rangle - \beta \langle E^{(1)} E^{(1)} \rangle_L, 
\end{eqnarray}
where subscript ``$L$'' means that only diagrammatically linked contributions are to be retained, and the last term (carrying a $\beta$ multiplier) is the anomalous-diagram term.
This is still far from the whole second-order internal energy [Eq.\ (74) of Ref.\ \onlinecite{HirataJhaJCP2020}], but the presence of the last anomalous-diagram term will make clear
that this ansatz is unworkable. 

The zeroth-order thermal averages in the above expression can be evaluated\cite{HirataJhaJCP2020,Hirata2021} as
\begin{eqnarray}
U^{(2)} 
&=&
\sum_{p,q}^{\text{denom.}\neq0} \frac{F_{qp} F_{pq} }{\epsilon_p^{(0)} - \epsilon_q^{(0)}} f_p^- f_q^+ \nonumber\\ 
&& -{\beta} \sum_{p,q}^{\text{denom.}=0} {F_{qp}  F_{pq} f_p^- f_q^+ } \nonumber\\ 
&& + 
\frac{1}{4} \sum_{p,q,r,s}^{\text{denom.}\neq0} \frac{\langle pq||rs \rangle \langle rs || pq \rangle}{\epsilon_p^{(0)} + \epsilon_q^{(0)} - \epsilon_r^{(0)} - \epsilon_s^{(0)}} f_p^-f_q^-f_r^+f_s^+
\nonumber\\
&& - \frac{\beta}{4} \sum_{p,q,r,s}^{\text{denom.}=0} {\langle pq||rs \rangle \langle rs || pq \rangle}f_p^-f_q^-f_r^+f_s^+, \label{app:U2}
\end{eqnarray}
whose diagrammatic representation is given in Fig.\ \ref{fig:U2}. 

The second and fourth terms (carrying a $\beta$ multiplier but no denominator) 
are anomalous-diagram terms,\cite{kohn} causing the low-temperature breakdown known as the Kohn--Luttinger nonconvergence.\cite{kohn,luttingerward,Hirata_KL2021,Hirata_KL2022}
Take the second term for example. The ``denom.=0'' restricts the summation to over those combinations of indexes whose fictitious denominator $\epsilon_p^{(0)} - \epsilon_q^{(0)}$ is zero.
One such combination is $p=q$. As $T$ is lowered to zero (i.e., $\beta \to \infty$),\cite{kohn}
\begin{eqnarray}
\lim_{T \to 0} \beta f_p^- f_p^+ = \delta\left( \mu - \epsilon_p^{(0)} \right), \label{delta}
\end{eqnarray}
where the right-hand side is Dirac's delta function, which is divergent when the chemical potential $\mu$ coincides with one of the orbital energies.\cite{kohn} 
Therefore, a Kohn--Luttinger nonconvergence occurs when $F_{pp} \neq 0$ and at the same time, the reference wave function is degenerate (i.e., the HOMO and LUMO 
have the same energy as $\mu$). While the divergence of this term can be avoided 
by adopting a HF reference wherein $F_{pq} = 0$,\cite{luttingerward} the fourth term remains divergent 
for $p=r$ and $q=s$ in a degenerate reference.\cite{Hirata_KL2021,Hirata_KL2022} 
It should be recalled that the second-order correction to the ground-state energy is always finite regardless of whether the zeroth-order ground state 
is degenerate\cite{Hirschfelder} or nondegenerate.\cite{shavitt} Hence, $U^{(2)}$ is not convergent at the correct zero-temperature limit, which is always finite. 

\begin{figure}
  \includegraphics[scale=0.4]{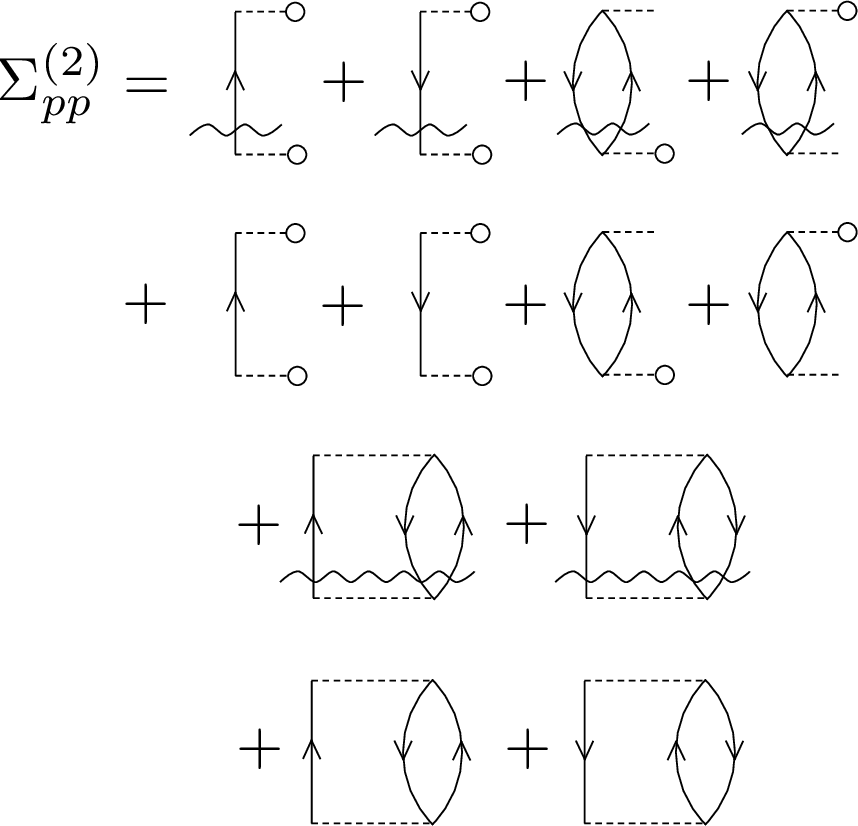}
\caption{Second-order thermal Dyson self-energy diagrams of Eq.\ (\ref{app:Sigma2}), which include anomalous diagrams with no resolvent lines.
They are obtained by deleting an edge in each of the internal energy diagrams of Fig.\ \ref{fig:U2}.}
\label{fig:Sigma2_e0_KL}
\end{figure}

As per Eq.\ (\ref{selfenergy}), the corresponding thermal self-energy is obtained by taking the derivative of $U^{(2)}$ with respect to $f_p^-$, leading to
\begin{eqnarray}
\Sigma_{pp}^{(2)} &\equiv& \frac{\partial U^{(2)}}{\partial f_p^- } \nonumber\\
&=&
\sum_{q}^{\text{denom.}\neq0} \frac{ F_{qp}F_{pq} }{\epsilon_p^{(0)} - \epsilon_q^{(0)}} f_q^+ 
 - \sum_{q}^{\text{denom.}\neq0} \frac{F_{pq}F_{qp} }{\epsilon_q^{(0)} - \epsilon_p^{(0)}} f_q^- \nonumber\\ 
&& + \sum_{q,r}^{\text{denom.}\neq0} \frac{\langle qp || rp \rangle F_{rq} }{\epsilon_r^{(0)} - \epsilon_q^{(0)}} f_r^- f_q^+ \nonumber\\
&& + \sum_{q,r}^{\text{denom.}\neq0} \frac{ F_{qr} \langle rp || qp \rangle }{\epsilon_r^{(0)} - \epsilon_q^{(0)}} f_r^- f_q^+ \nonumber\\
&& -{\beta} \sum_{q}^{\text{denom.}=0} F_{qp}F_{pq}  f_q^+  + {\beta} \sum_{q}^{\text{denom.}=0} F_{pq}F_{qp}  f_q^-  \nonumber\\ 
&& -{\beta} \sum_{q,r}^{\text{denom.}=0} {\langle qp || rp \rangle  F_{rq} f_r^- f_q^+ } \nonumber\\ 
&& -{\beta} \sum_{q,r}^{\text{denom.}=0} {F_{qr} \langle rp || qp \rangle  f_r^- f_q^+ } \nonumber\\ 
&& +  \frac{1}{2} \sum_{q,r,s}^{\text{denom.}\neq0} \frac{\langle pq||rs \rangle \langle rs || pq \rangle}{\epsilon_p^{(0)} + \epsilon_q^{(0)} - \epsilon_r^{(0)} - \epsilon_s^{(0)}}  f_q^-f_r^+f_s^+
\nonumber\\
&& - \frac{1}{2} \sum_{q,r,s}^{\text{denom.}\neq0} \frac{\langle rs || pq \rangle \langle pq||rs \rangle}{\epsilon_r^{(0)} + \epsilon_s^{(0)} - \epsilon_p^{(0)} - \epsilon_q^{(0)}}  f_q^+f_r^-f_s^-
\nonumber\\
&& - \frac{\beta}{2} \sum_{q,r,s}^{\text{denom.}=0} {\langle pq||rs \rangle \langle rs || pq \rangle} f_q^-f_r^+f_s^+
\nonumber\\
&& + \frac{\beta}{2} \sum_{q,r,s}^{\text{denom.}=0} {\langle rs || pq \rangle \langle pq||rs \rangle } f_q^+f_r^-f_s^-.
 \label{app:Sigma2}
\end{eqnarray}
The corresponding diagrams are shown in Fig.\ \ref{fig:Sigma2_e0_KL}, which are obtained by opening each of the $U^{(2)}$ diagrams in Fig.\ \ref{fig:U2} by deleting an edge. 
Note the presence of numerous anomalous diagrams.

This formula suffers from a more pervasive kind of the Kohn--Luttinger nonconvergence problem. 
Take the last term for example. It sums over a set of indexes whose fictitious denominator $\epsilon_r^{(0)} + \epsilon_s^{(0)} - \epsilon_p^{(0)} - \epsilon_q^{(0)}$ is zero.
It includes the $p=r$ and $q=s$ case, whose summand diverges as $T \to 0$ ($\beta \to \infty$)
for any reference---degenerate or nondegenerate---because a $f_p^+$ factor is no longer there to create a delta function [Eq.\ (\ref{delta})] which diverges only when $\mu = \epsilon_p^{(0)}$. 
In other words, a finite-temperature generalization of the self-energy containing anomalous-diagram terms 
is always divergent as $T \to 0$
and will be uselessly erroneous at low $T$. 

This is not surprising because in thermodynamics all states with any number of electrons are summed over in the partition function and they include 
numerous exactly degenerate zeroth-order states. This is why we purposefully neglected anomalous-diagram terms when defining $U^\text{QP(2)}$. Generally, 
it is inappropriate to include them in the internal energy or grand potential, when their (closed) diagrams will be opened to form some kind of potentials. 

This  bodes ill for the Luttinger--Ward functional\cite{luttingerward,BaymKadanoff1961,Baym_selfconsistent,Kozik2015,Rossi2015,Zgid2016,Gunnarsson2017,Lin2018} 
because opening a diagram of this functional may also result in thermal self-energy and Green's function diagrams that  suffer from the same severe low-temperature breakdown. 

\section{Feynman--Dyson nonconvergence\label{sec:appendix2}}

We reported\cite{Hirata_PRA2024} pervasive divergences of a Feynman--Dyson perturbation expansion of the self-energy in many frequency domains. 
Therefore, apart from the Kohn--Luttinger-type nonconvergence of the {\it thermal} self-energy caused by anomalous-diagram terms, the {\it zero-temperature} 
self-energy will already be divergent or at least excessively erroneous for low- and high-lying states. This is alleviated (or hidden from view) by the frequency-independent approximation inherent in the ansatz
of thermal quasi-particle theory, but in a solid, whose self-energy forms continuous energy bands, we can no longer hope to be able to avoid this deep-rooted pathology.\cite{Hirata_PRA2024}

In a solid-state implementation of thermal QP(2) theory (which is underway in our laboratory), therefore,
we will have to introduce active orbitals (or active energy bands) and evaluate only some portions of $\langle E^{(2)} \rangle$ and $\Sigma_{pp}^{(2)}$
involving  them. These active orbitals are the ones whose zeroth-order energies fall within the so-called ``central overlapping bracket'' where 
the Feynman--Dyson perturbation expansion is guaranteed to have a nonzero radius of convergence.\cite{Hirata_PRA2024}
Comparative performance of this method for solids will be presented in the future. 

\bibliography{library.bib}

\end{document}